# Multiferroicity and magnetoelastic coupling in α-$Mn_2O_3$: A binary perovskite

Mohit Chandra[1], Satish Yadav[1], R. J. Choudhary[1], R. Rawat[1], A. K. Sinha[2], Marie-Bernadette Lepetit[3,4] and Kiran Singh[1*]

[1]UGC-DAE Consortium for Scientific Research, University Campus, Khandwa Road, Indore - 452001, India

[2] HXAL, Synchrotrons Utilization Section, RRCAT, Indore 452013, India

[3] Institut Néel, CNRS UPR 2940, 25 av. des Martyrs, 38042 Grenoble, France

[4]Institut Laue Langevin, 72 av. des Martyrs, 38042 Grenoble, France

Abstract

Multiferroics where at least two primary ferroic orders are present and coupled in a single system constitute an important class of materials. They attracted special consideration as they present both intriguing fundamental physics problems and technological importance for potential multifunctional devices. Here, we present the evidence of multiferroicity and magnetoelectric (ME) coupling in α-$Mn_2O_3$; a unique binary perovskite. Corresponding to the antiferromagnetic (AFM) ordering around 80K, a clear frequency independent transition is observed in the dielectric permittivity. We showed that electric polarization emerges near AFM regime that can be modulated with magnetic field. The detailed structural analysis using synchrotron radiation X-ray diffraction demonstrates the increase in structural distortion with decreasing temperature, as well as changes in the unit cell parameters and bond lengths across the ferroelectric and magnetic ordering temperatures. This observation of multiferroicity and magnetoelastic coupling in α-$Mn_2O_3$ provides insights for the exploration of ME coupling in related materials.

*Corresponding author:    Kiran Singh

kpatyal@gmail.com

The search for new multiferroics material increased enormously in recent years due to their fascinating interplay between magnetic and electric order parameters [1-4]. The magnetoelectric (ME) coupling was first observed in $Cr_2O_3$ in 1960 [5-8]. It gained a strong impetus in the recent years after the discovery of very large ME coupling in the $TbMnO_3$, and $TbMn_2O_5$ compounds [9-10]. The exploration for new ME materials has been mainly sustained by the conjunction of the physical challenge in the understanding of their unique properties and by the prospect of technological applications in spintronics. It has resulted in a remarkable expansion of the field [2, 11-21] and the notice that the transition metal oxides are the most suitable family where new ME materials can be found.

Manganese present various stable oxidation states in its oxides that exhibit many fascinating properties, as for instance colossal magneto-resistance with a record over 14 orders of magnitude in resistivity change under magnetic field [22], and giant magnetocapacitance [23]. Among these oxides, $Mn_2O_3$ has attracted lot of attention due to its various applications in energy and environmental fields [24-26]. The chemical formula of the $Mn_2O_3$ compound can be rewritten as $MnMnO_3$ which is analogous to a $ABO_3$ perovskite, with a Mn ion both in the A and B sites. In perovskites, the magnetic and electronic properties are mainly ascribed by B-O-B interactions. The replacement of the A site cation by a transition metal leads to a new class of compounds, where magnetic and transport properties can be controlled both by the A-O-A and A-O-B interactions, in addition to the B-O-B ones. $Mn_2O_3$ is a unique material, not only as it is the only binary oxide which crystallize in a perovskite structure, but also because the Mn cations occupying A and B sites have the same charge states [27]. Among trivalent ion sesquioxides of Ti, V, Cr, Fe and Ga; $Mn_2O_3$ is the only one that does not crystallize in a corundum-type structure. Initially, $\alpha$-$Mn_2O_3$ was considered to be crystallized in cubic symmetry with space group *Ia*-3 (*bixbyite*) [28,29]. Subsequently, Geller et al. reported that it rather presents an orthorhombic (*Pcab*) structure at room temperature (RT) [28] and undergoes cubic transition around 308 K [30, 31]. We will see in this work that this space group should be further probed, at least at low temperature.

Magnetically $\alpha$-$Mn_2O_3$ is antiferromagnetic (AFM) with an ordering temperature around ~80 K [32,33]. The magnetic structure of this material is very complex and has not been solved unequivocally. Initially only one transition was observed around 80 K, but later a second transition was detected around 25 K [30] and suggested to be *first-order*. The powder neutron diffraction (PND) results show that the magnetic unit cell is equivalent to crystal unit cell [34-37]. Regulski et al. [36] suggested a collinear AFM structure, however, recently, Cockayne et al. [37] suggested the possibility of collinear and non-collinear magnetic arrangements.

At RT, the cubic to orthorhombic distortion is very small and it is difficult to distinguish the two symmetry groups using laboratory based diffraction experiments. It is thus essential to have the resolution of synchrotron X-ray diffraction (SXRD) to see the distortion or possible additional peaks in the diffraction pattern. Stoichiometrically $Mn_2O_3$ resembles the *first* known ME material, namely $Cr_2O_3$, but has not been explored for its dielectric and ME properties. Here, we report the evidence of electric polarization and magnetoelectric coupling in α-$Mn_2O_3$ using temperature dependent magnetic, dielectric, polarization and SXRD measurements.

**Results**

**Room temperature synchrotron based x-ray diffraction.** The RT laboratory based XRD and neutron powder diffraction (NPD) of the prepared samples are reported earlier [38] and confirm the phase purity of the prepared samples. It is however rather difficult to state conclusively on the exact crystal symmetry from those results. Indeed, the diffraction pattern is equally fitted using the cubic (*Ia*-3) or orthorhombic (*Pcab*) space groups (not shown here). This ambiguity is due to the weakness of the distortions in the crystal structure at RT. Hence, we performed SXRD measurements. The RT Rietveld refined SXRD pattern of α-$Mn_2O_3$ using *Pcab* space group is shown in Fig. 1a. The refined reliability parameters are given inside Fig. 1a. The refined lattice parameters a=9.4016(5), b=9.4052(4) and c=9.3985(2) are consistent with earlier reports [39]. We also fitted the data using cubic symmetry. For the sake of clarity, Bragg reflections (vertical lines in Fig. 1a) for cubic as well as orthorhombic symmetries are denoted with vertical pink and green color lines, respectively. From Fig. 1a, it is clear that there are many additional peaks of small intensity, that are not allowed in the cubic space group. For example, at 2θ~24.30° (see inset of Fig. 1a), there is a weak peak denoted as (034), and $h^2+k^2+l^2= 25$, allowed in the *Pcab* space group that is forbidden in *Ia*-3 by the extinction rules. The presence of *0kl* (*l*=2n) planes is supporting the orthorhombic space group [39]. We also observed peaks at d=3.134 and 1.167 Å which are consistent with Geller et al. [28]. We were not able to resolve the splitting of the $h^2+k^2+l^2 = 64$ {008} and 66 {811} peaks at RT, but nevertheless a peak asymmetry can be observed even at RT. We can thus conclude without ambiguity on the orthorhombic symmetry of $Mn_2O_3$ at RT.

The schematic view of the $Mn_2O_3$ crystal structure is pictured in Fig. 1b. It consists of 35 Mn atoms per unit cell in octahedral environment. All the octahedra are distorted, even at RT. The nearest neighbor octahedra are bond-sharing while each oxygen atom is surrounded by four Mn atoms.

**Magnetic susceptibility.** The temperature dependent dc magnetization results are presented in Fig. 2. Two clear magnetic transitions can be seen in Fig. 2a; one at 80 K and another at 25 K, in agreement with earlier reports [30]. The later transition is merely observed in samples prepared by solid state reaction [30]. The fact that we observe a clear transition in our polycrystalline samples corroborates the good quality of our samples. As mentioned earlier, the 25 K transition was suggested to be of *first-order* nature [30] but no clear evidence was given. To further explore the nature of this transition, we performed zero field cooled (zfc) warming, field cooled cooling (fcc) and field cooled warming (fcw) magnetic measurements at 500 Oe. A clear hysteretic behavior is observed in fcc and fcw curves between 20 and 30 K (see inset of Fig. 2a). This observation clearly establishes the *first-order* nature of the 25 K transition [30]. The inverse magnetic susceptibility follows a Curie-Weiss law between 200 K and 300 K (not shown here). The calculated effective paramagnetic moment ($\mu_{eff}$) per $Mn^{3+}$ ion is found to be 4.87 $\mu_B$, which is approximately equal to the theoretical spin only value of 4.90 $\mu_B$ for a high spin $Mn^{3+}$ ion. This value is consistent with earlier work [32]. The Weiss constant ($\theta p$) is found to be -25.59 K, its negative value confirming the predominating AFM interactions. The frustration index ($f$ ($|\theta_c/T_N|$) is 1.4 which suggest geometrical frustration and endorses the earlier interpretation of frustrated magnetic interactions in this material [37].

The first derivative of the fcw magnetic susceptibility is shown in Fig. 2b. It evidences a sharp transition around both 80 K and 25 K. Moreover, there is a broad feature around 50 K, and a tail like feature above the sharp peak at 80 K that ends around 110 K. This tail should be put into perspective with the remanence of the (100) magnetic Bragg peak up to over 100 K, seen in neutron scattering by Regulski et al. [36]. These effects can be interpreted as the sign of a magnetic order parameter onset around 110 K. Whether this onset is associated with a true long range order or a short range one is an open discussion. Indeed, on one hand, heat capacity measurements do not exhibit any sign of a phase transition in the 105-115 K temperature range, while clear anomalies can be seen at the 80 K [40]; and on another hand Regulski et al. [36] argue against a short-range order as the width of the (100) Bragg peak is the same below and above the 80 K magnetic transition. Finally, the isothermal magnetization M(H) at 2 K is shown up to 50 kOe field in the inset of Fig. 2b; it perfectly illustrates the linear behavior expected from an antiferromagnet.

**Dielectric permittivity.** The α-$Mn_2O_3$ is a semiconducting material with high resistivity even at RT, which is the prerequisite for dielectric studies. The temperature and frequency dependent complex dielectric behavior is measured from 5-300 K. The temperature dependent real part of dielectric permittivity at 100 kHz during warming (1 K/min) is presented in Fig. 3a. A clear transition is observed at 80 K which coincides with the AFM transition and infer a ME coupling in this material. This transition is frequency independent (not shown here). The change in the magnitude of dielectric permittivity is very small below 120 K. The tan$\delta$ value is also very small below 120 K (<$10^{-3}$, not shown here) which infer the insulating behavior of the studied sample. The sharp change in the dielectric permittivity at 80 K is similar to what is observed in other well-known ME materials e.g. $YMnO_3$, $LuMnO_3$, MnO and $BaMnF_4$ etc. [41-43]. Such a decrease in the dielectric constant at a magnetic transition is sometimes (partly) attributed to the geometrical frustration, as in some $RMnO_3$ [41]. No visible change in the dielectric permittivity was however detected around 25 K.

To see the effect of a magnetic field on the transition temperature, we measured the dielectric behavior under different magnetic fields (0- 80 kOe). The data at some selected fields are presented in the inset of Fig. 3a. As one can see there is no shift in the transition temperature (80 K) seen in the dielectric permittivity, nor any field induced transition at 25 K. This invariance of $T_N$ suggests the robustness of the AFM ordering at least up to 80 kOe. The derivative of the real part of the dielectric permittivity with respect to temperature, measured at zero magnetic field and 100 kHz, is presented in Fig. 3b. As expected, it exhibits a clear peak at 80 K, whereas a feeble change in slope can be noticed around 25-30 K. Importantly, Fig. 3b also evidences a tail above the 80 K transition that lasts up to 110 K. The isothermal magnetodielectric (MD) behavior at 10 K is shown in inset of Fig. 3b where MD is defined as $(\varepsilon'_H - \varepsilon'_{H=0})/\varepsilon'_{H=0}$ The magnitude of MD is rather small but in agreement with ME materials [44, 45]. The comparison of the M(H) and MD results confirm the existence of a ME coupling in this material.

**Electric polarization.** To explore the ferroelectric behavior of this material, temperature dependent remnant electric polarization (*P*) is measured as described in the method section. Fig. 4 illustrates the variation of *P* with temperature. From this figure, it is clear that *P* decreases sharply above 80 K and becomes *T* independent only around 110 K. The *P* reversal is observed by reversing the polarity of the poling electric field, thus confirming the ferroelectric behavior of α-$Mn_2O_3$. The remnant *P* is ~40 $\mu C/m^2$ at 10 K, comparable in amplitude with other multiferroics including $Cr_2O_3$ [8, 9]. A similar experiment is performed in the presence of a 5 kOe magnetic field and a clear effect of the field is observed below AFM ordering, confirming the ME coupling in this material. Additionally, a convex downward feature near 15 K and a peak like behavior at 24 K are observed, consistent with magnetization results (Fig. 2b). Finally, there is an additional feature around 50 K i.e. *P* does not vary smoothly with increasing temperature but has concave upwards slope around 50 K. This change occurs at the same temperature where the crossing in intensity between the (110) and the (111) magnetic peaks is observed in neutron scattering [35]. Similarly, it is associated with the anomalies seen in the 1% Fe doped Mossbauer results [30] as well as our magnetization data.

**Temperature dependent structural evolution.** In order to understand the structural correlations between the magnetic and dielectric properties, we performed *T* dependent SXRD down to 6 K. Our analysis reveals that the crystal structure remains orthorhombic down to 6 K. However, the distortion increases with decreasing temperature, and the splitting of some peaks can clearly be seen at higher 2θ angles. The RT SXRD pattern is very well refined within the *Pcab* space group. This group belongs to the *mmm* point group (centrosymmetric) and does not allow for an electric polarization. Our experimental results however demonstrate the emergence of an electric polarization below 110 K, thus suggesting a lower symmetry, non-centrosymmetric space group at least below 110 K. At this point one should remember two things. First that the polarization measurements using pyroelectric method see only changes in the polarization and not absolute values. Second that the crystal structure of the $RMn_2O_5$ compounds, a benchmark example of ME materials, was believed to be centrosymmetric for a long time [10], but was recently found non-centrosymmetric even at RT. Such a discovery required a revision of the origin of the polarization in the magnetically ordered state as a spin-enhanced process rather than a spin-induced one [46]. We thus performed a symmetry analysis and looked for *Pcab* subgroups allowing an electric polarization. The *Pca*$2_1$ group is the only polar maximal subgroup of *Pcab* loosing only one generator (index 2). We thus analyzed the temperature dependent SXRD patterns using both space group; *Pcab* as well as *Pca*$2_1$.

In $Pca2_1$ the 8b and 8c Wickoff positions split into 4a + 4a, that is the $Mn_3$, $Mn_4$, $Mn_5$ and all oxygens split into double setting. The calculated diffraction in $Pca2_1$ allows additional Bragg reflection (130) around $2\theta=15.45°$ but we did not observe any peak intensity (whatever the temperature) at this $2\theta$ value within our experimental resolution. The refined unit cell parameters remained the same in both space groups.

The Rietveld refined SXRD patterns using $Pca2_1$ at some selected temperatures are shown in Fig. 5 (a-d). They reveal the good agreement between calculated and observed patterns. To show the distinction/similarity in our refined pattern using both space groups, we presented the refinement at 80 K using $Pca2_1$ and $Pcab$ in Fig. 5b, c. We cannot observe any significant difference in the refinements and unit cell parameters. The main intensity peak does not undergo any splitting down to lowest temperature; in fact, the maximum changes are observed at higher $2\theta$ range. Hence, for visual inspection, the diffraction pattern is normalized with the main peak intensity. The normalized intensity of the selected $2\theta$ range, mainly related to {800} and {811} family, is presented in Fig. 5e at selected temperatures. In addition to the $h^2+k^2+l^2=64$ and 66 there are peaks with $h^2+k^2+l^2=65$. It is worth mentioning here that the x and y scales for all patterns in Fig. 5e are the same. The Fig. 5e distinctly demonstrates the change/distortion in lattice with decreasing temperature. For instance, at 300 K, there is no clear splitting in the above mentioned peaks but some asymmetric features, whereas splitting enhances progressively with decreasing temperature. The vertical red arrow in the 240 K data is shown as a guide to the eyes to demonstrate that this height increases with decreasing temperature. One can clearly notice the change in the peak intensities across magnetic transitions; e.g. in the $2\theta$ range 39.60° to 40.10°, there was only two visible peaks up to 140 K, whereas one additional peak starts appearing in the middle of these two peaks below 110 K (see peaks for 100-60 K). With further decrease in temperature (50-6 K), the intensity of (800) peak becomes less visible because the relative intensity of these peaks increases, see especially at 20 K. These results infer the change in lattice distortion across the magnetic transitions and speaks in favor of a magnetoelastic coupling.

The evolution of the unit cell parameters with temperature is presented in Fig. 6a, b. Near RT the three lattice parameters have similar values, as expected when one is close to a cubic to orthorhombic phase transition ($T_c$=308 K [30,31]). They however differ significantly when the temperature decreases (see Fig. 6a). The distortion parameter (b/c) increases with decreasing temperature (Fig. 6b) with a maximum plateau between 80 K and 110 K, then it starts to decrease, consistently with earlier reports [37]. A small but clear change in all the lattice parameters as well as in the unit cell volume is observed at the 80 K and 25 K magnetic transitions.

The temperature evolution of selected Mn bond lengths obtained from refined patterns are presented in Fig. 6c-e. The temperature evolution is different for the different Mn-O bonds. For example, the $Mn_{4(1)}$-$O_{5(2)}$ bond length decreases with increasing temperature (Fig. 6c) whereas $Mn_{3(1)}$-$O_{5(2)}$ bond length increases with increasing temperature (Fig. 6d). A clear change in these bond lengths can be seen around 80 K with a tail like feature up to 110 K. The bond lengths of $Mn_{3(1)}$-$O_{4(1)}$ is displayed in Fig. 6e. These bond lengths show very clear anomalies at ~30 and ~80 K, that coincide with magnetic transitions. In addition, these bond lengths show a broad peak like feature between 80 K and 110 K. A schematic view of the four Mn atoms surrounding the $O_{5(2)}$ is shown in the inset of Fig. 6c. The details of some bond lengths and bond angles at selected temperatures are given in Table I.

**Discussion and conclusion.** Our SXRD results confirm that α-$Mn_2O_3$ crystallize in an orthorhombic symmetry at RT, however, these data do not present any direct experimental evidence favoring the polar $Pca2_1$ space group over the non-polar $Pcab$ one, while our polarization results clearly imply a symmetry lowering from $Pcab$ to a polar subgroup. Moreover, the refined structures in the two groups are associated with different temperature evolution of the Mn-O bond lengths. Indeed, while the $Pca2_1$-refined structure exhibit clear signs of the three anomalies (the two magnetic transitions at 25 K and 80 K and the polar/magnetoelectric transition at 110 K), these signatures are not as obvious when the structure is refined in the $Pcab$ space group (see Fig. S1 in the supplementary information). These suggest that $Pca2_1$ could be the correct space group.

The dielectric/polarization and magnetic measurements clearly reveals the magnetoelectric nature of α-$Mn_2O_3$ through various signs of coupling between the magnetic and polar orders. The electric polarization emerges around 110 K that is at higher temperature than the main transition observed in dielectric and magnetization results (~ 80 K). However, a clear anomaly in the magnetic susceptibility derivative is observed at the onset of the polar order, coherent with the earlier report of the extinction of one of the magnetic peaks (100) at only ~110K. Despite our efforts the exact nature of this transition remains unclear. Indeed, Regulski et al. [36] pointed out that the width of the (100) magnetic peak below and above 80 K is temperature independent and hence cannot be related to short range magnetic ordering. They rather suggested the existence of a structural distortion at ~110 K. While such a transition would fit very well with our polarization results, we were unable to see any sign of transition in X-Ray diffraction data. If one rather supposes the absence of a structural transition at 110 K, then one should assume that there is no onset of a polar order at 110 K, but rather the onset of a temperature dependent behavior of the polarization as in the $RMn_2O_5$ family [46]. In this case the space group should be *Pca*$2_1$ in the whole temperature range, up to RT. This hypothesis is supported by the behavior of the Mn-O bond-length as a function of temperature, when the structure is refined in this group, clear anomalies allow to identify all transitions, while this is not the case when the data are refined in the *Pcab* group. In this hypothesis the observed electric polarization under 110 K could originate from spin components due to magnetoelectric coupling like in $Dy_2Mn_2O_5$ [46]. The one to one correspondence of the features in the magnetic, dielectric, electric polarization and structural properties demonstrate the coupling between these different order parameters in α-$Mn_2O_3$.

To summarize, we performed a comprehensive study on α-$Mn_2O_3$ to explore its multiferroicity and magnetoelastic coupling by investigating its structural, magnetic, dielectric and polarization behavior. The structural analysis confirmed the orthorhombic structure at room temperature that remains *invariant* down to 6 K. However, the lattice distortion (b/c) increases with decreasing temperature and shows a maximum around 100 K. Clear changes in lattice parameters observed across the magnetic transitions support the importance of the magnetoelastic coupling. Associated with the magnetic transitions, clear anomalies in the dielectric permittivity are observed. Similarly, the electric polarization can be tuned by a magnetic field, showing the existence of a ME coupling in this material. The emergence of a polarization suggest that the structural space group could be *Pca*$2_1$, a subgroup of *Pcab*. Our detailed studies thus unambiguously established multiferroicity and magnetoelastic coupling in a unique binary perovskite material α-$Mn_2O_3$ above liquid nitrogen temperature and suggest the search for such coupling in related materials.

**Acknowledgements**

The authors would also like to thank Prof. E.V. Sampathkumaran and Dr. Vengadesh for their help in performing initial dielectric and magnetic measurements. Authors are also thankful to Dr. Archana Sagdeo and Mr. M. N. Singh for their help during low temperature SXRD measurements.

**Methods**

The polycrystalline samples of α-$Mn_2O_3$ were prepared through standard solid state reaction method using $MnO_2$ (99.99%) as reported earlier [38]. The temperature dependent synchrotron X-ray diffraction (SXRD) measurements were performed at angle dispersive X-ray diffraction beamline (BL-12) [47] at Indus-2, RRCAT Indore ($\lambda$=0.8002 Å). The data conversion is made using the Fit2D software [48]. The dc magnetization measurements were done on a 7 T Quantum Design SQUID-VSM. The temperature and magnetic field dependent complex dielectric measurements were performed using a home-made insert coupled with a 8 T Oxford magnet and a Keysight E4980A LCR meter. A computerized based data acquisition software was developed using the LabVIEW platform. A parallel plate capacitor geometry with silver paste as electrodes was used to perform these measurements. The temperature dependent dielectric measurements were performed at different frequencies (1 kHz–100 kHz) and different magnetic fields (0, 5, 10 and 80 kOe). Isothermal magnetodielectric measurements were performed at selected temperatures up to 70 kOe, with a field sweep rate of 5 kOe/min. The remnant electric polarization ($P$) was measured using a Keithley 6517B electrometer. A poling electric field of ±310 kV/m was applied during cooling to align the electric dipoles, and removed at 8 K; then the sample was short circuited for 30 minutes to remove the extrinsic charges. After that charge *vs* time was recorded for one hour to remove the stray charge (if any) and then charge *vs* temperature is recorded during warming (1.5 K/min). The same procedure is followed to measure $P$ under magnetic field; however, a magnetic field (80 kOe) was applied at the lowest temperature.

**Data availability.** All data supporting this study are available in the paper and its Supplementary Information. Extra data can be obtained from the corresponding authors on request.

**Competing interests:** The authors declare no competing financial interests.

**Author contributions.** K. S. conceived the problem and supervised the whole project. M. C. prepared the sample. M. C., S. Y. and K. S. designed the sample holder for performing dielectric, magnetodielectric and polarization measurements in 8 T Oxford magnet with the help of R.R. M. C., S. Y. and K. S. carried out experiments. M. C. analyzed the data with K. S. R.J.C. performed magnetization measurements and A.K.S. contributed in synchrotron based experiments. M.B.L. performed symmetry analysis. K.S. wrote the manuscript with the consultation from other authors. All authors contributed to the scientific discussion.

**Table Caption:**

**Table I.** Bond lengths and bond angles around oxygen ($O_5$) at selective temperatures across magnetic transition.

**Figure Captions:**

**Fig. 1. (a)** Reitveld refinement of room temperature (RT) SXRD data. The green and pink vertical lines represent Bragg's reflections for orthorhombic and cubic symmetry, respectively. Inset shows the zoomed view of SXRD to illustrate the orthorhombic peak.

**(b)** Schematic of crystal structure at RT.

**Fig. 2.** (a) Magnetic susceptibility (dc) *vs* temperature behavior at 500 Oe in different protocol; zero field cooled warming (zfc), field cooled cooling (fcc) and field cooled warming (fcw). The inset shows fcc and fcw curves near 25 K transition. (b) First derivative of fcw magnetization at 5 kOe and M(H) at 2 K.

**Fig. 3. (a)** Temperature variation of real part of dielectric permittivity at 100 kHz at selected magnetic fields (0 and 5 kOe) during warming (1K/min). Inset shows dielectric permittivity near $T_N$ at selective fields. **(b)** Derivative of dielectric permittivity measured without magnetic field at 100 kHz and inset shows magnetodielectric (MD) at 10 K. (c)

**Fig. 4.** Temperature dependent remnant electric polarization (*P*) with positive and negative poling electric fields. The blue curve represents *P* at 5 kOe magnetic field.

**Fig. 5. (a-d)** Rietveld refinement of SXRD patterns at some selected temperatures. **(e)** Normalized intensity of selected 2θ range to illustrate the splitting of {800} and {811} peaks at selected temperatures.

**Fig. 6.** Temperature dependent **(a)** lattice parameters and **(b)** lattice distortion **(b/c)** (left y-axis) and unit cell volume (right y-axis). Evolution of bond lengths as a function of temperature **(c)** $Mn_{4(1)}$-$O_{5(2)}$ **(d)** $Mn_{3(1)}$-$O_{5(2)}$, **(e)** $Mn_{3(1)}$-$O_{4(1)}$; inset in **(c)** shows a schematic representation of oxygen $O_{5(2)}$ with four Mn ions at 65 K.

**Table I**

| Bonds/ Angles | 105 K | 65 K | 6 K |
|---|---|---|---|
| **Bond Lengths [Å]** | | | |
| $Mn_2$-$O_{5(1)}$ | 2.0348 (6) | 1.9911 (7) | 1.9644(4) |
| $Mn_2$-$O_{5(2)}$ | 1.9993(7) | 1.9956(6) | 2.0389(6) |
| $Mn_{3(1)}$-$O_{5(2)}$ | 1.8686 (1) | 1.8545(4) | 1.8147(5) |
| $Mn_{3(2)}$-$O_{5(1)}$ | 1.9166(7) | 1.9272(2) | 1.9303(1) |
| $Mn_{4(1)}$-$O_{5(2)}$ | 1.9741(1) | 2.0218(7) | 2.0522(5) |
| $Mn_{4(2)}$-$O_{5(1)}$ | 1.9507(2) | 1.9757(1) | 1.9949(6) |
| $Mn_{5(1)}$-$O_{5(2)}$ | 2.2309(4) | 2.1990(3) | 2.1805(2) |
| $Mn_{5(2)}$-$O_{5(1)}$ | 2.1835(4) | 2.2067(1) | 2.23708 |
| **Bond Angles[deg.]** | | | |
| $Mn_{5(1)}$-$O_{5(2)}$-$Mn_{4(1)}$ | 95.2275 | 94.7489(2) | 93.8494 |
| $Mn_{5(1)}$-$O_{5(2)}$-$Mn_{3(1)}$ | 115.6564 | 116.5622 | 119.2851 |
| $Mn_{3(1)}$-$O_{5(2)}$-$Mn_{2(1)}$ | 136.0936 | 136.0947 | 135.812 |
| $Mn_{2(1)}$-$O_{5(2)}$-$Mn_{4(1)}$ | 101.9773 | 101.1528 | 98.3620 |

(a)

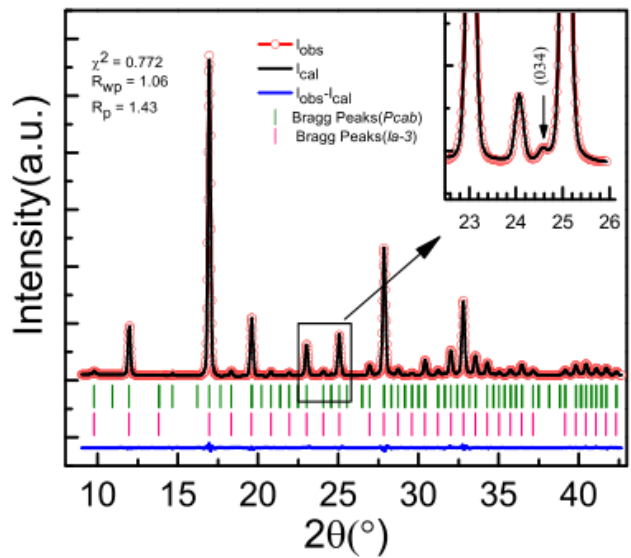

(b)

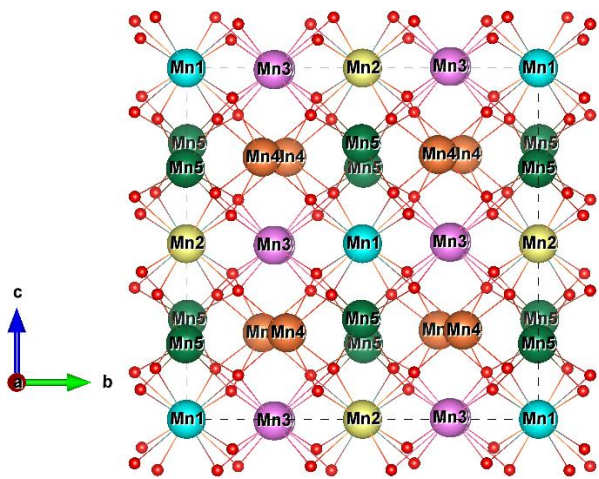

**Fig. 1**

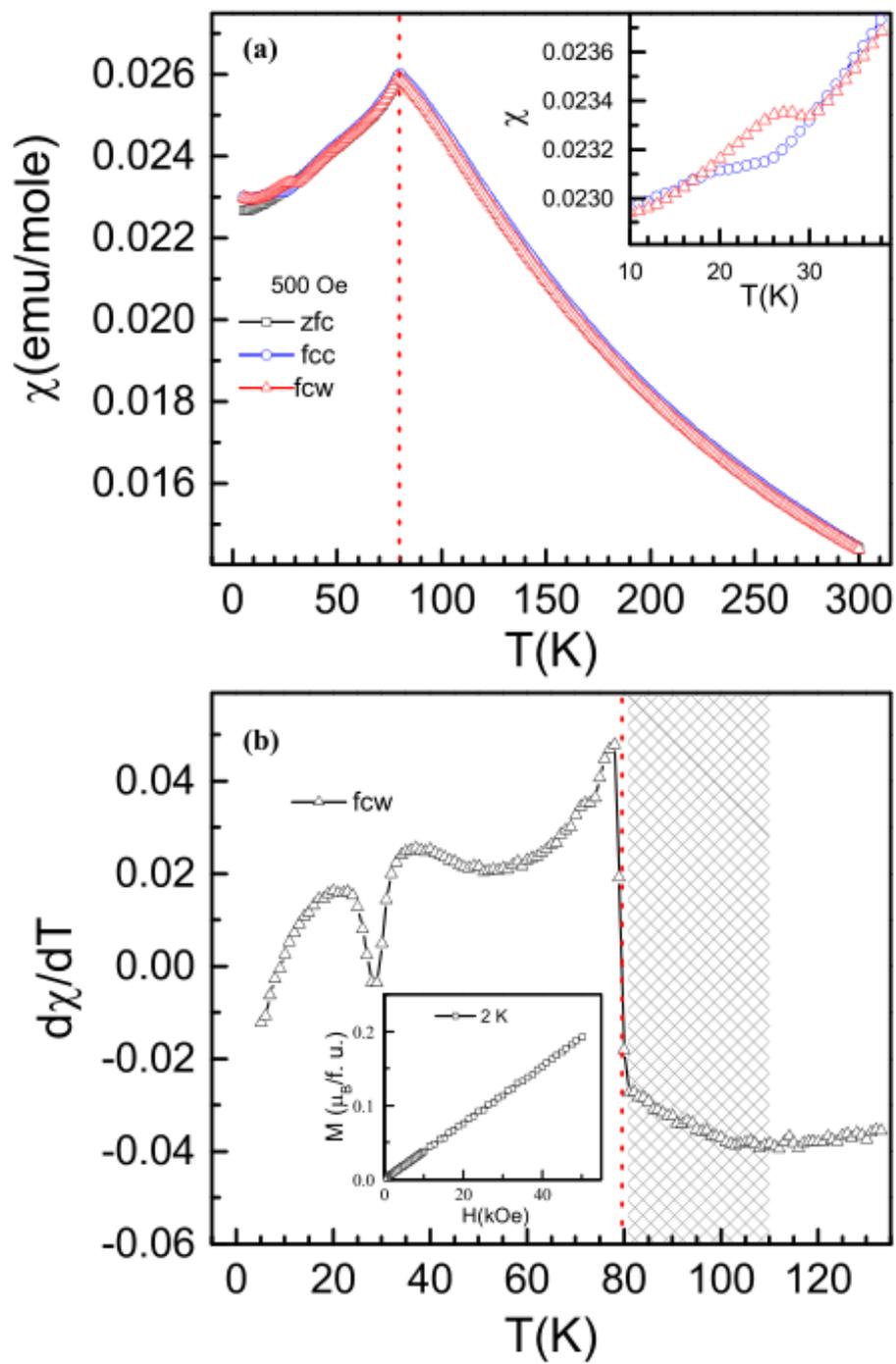

**Fig. 2**

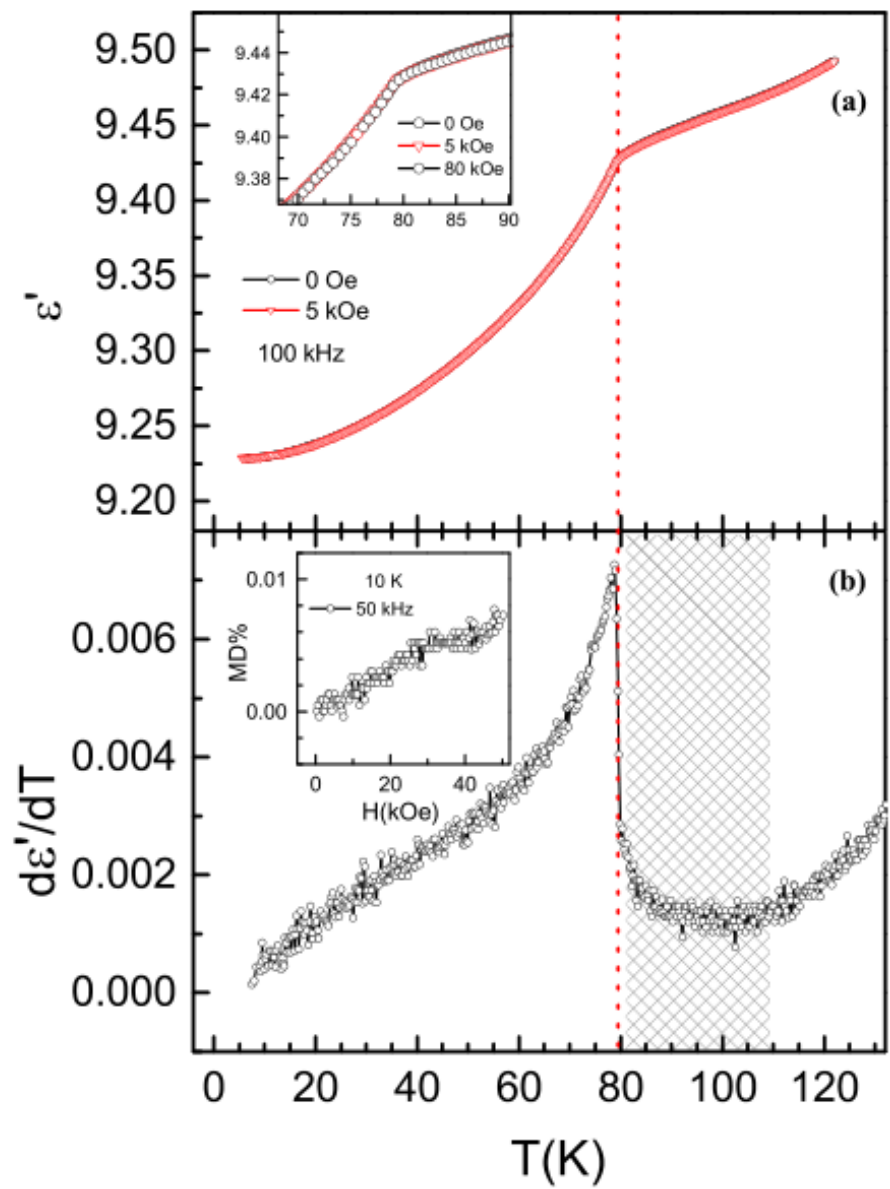

**Fig. 3**

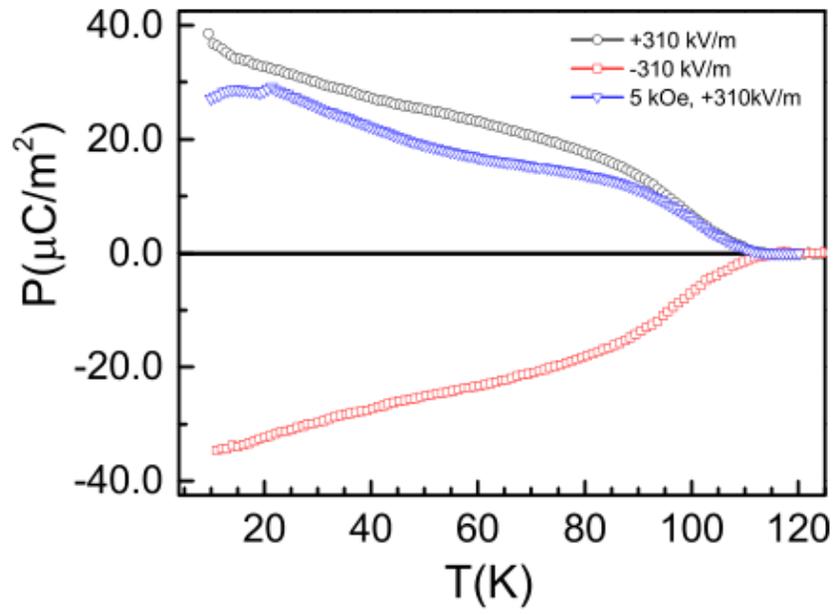

**Fig. 4**

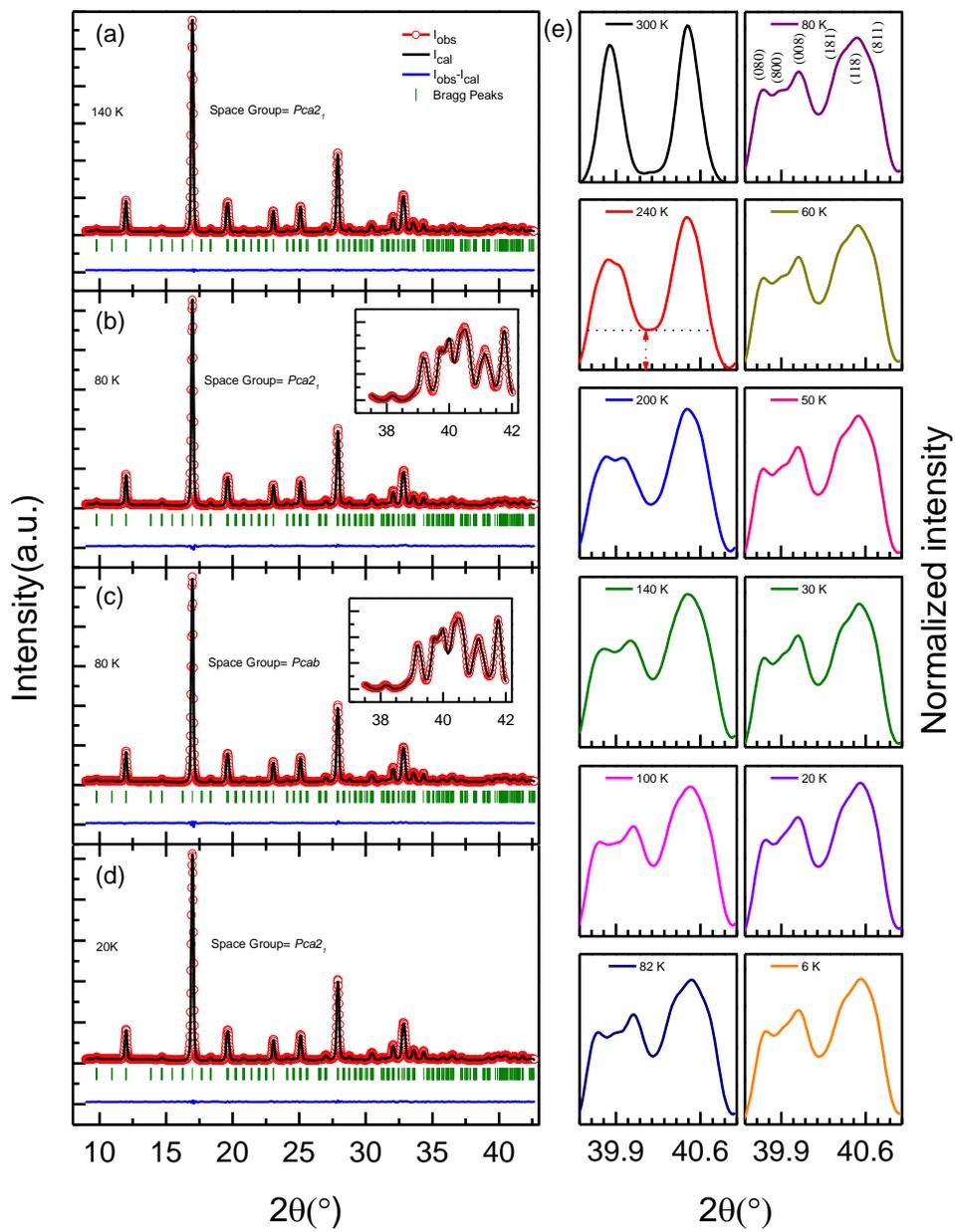

**Fig. 5**

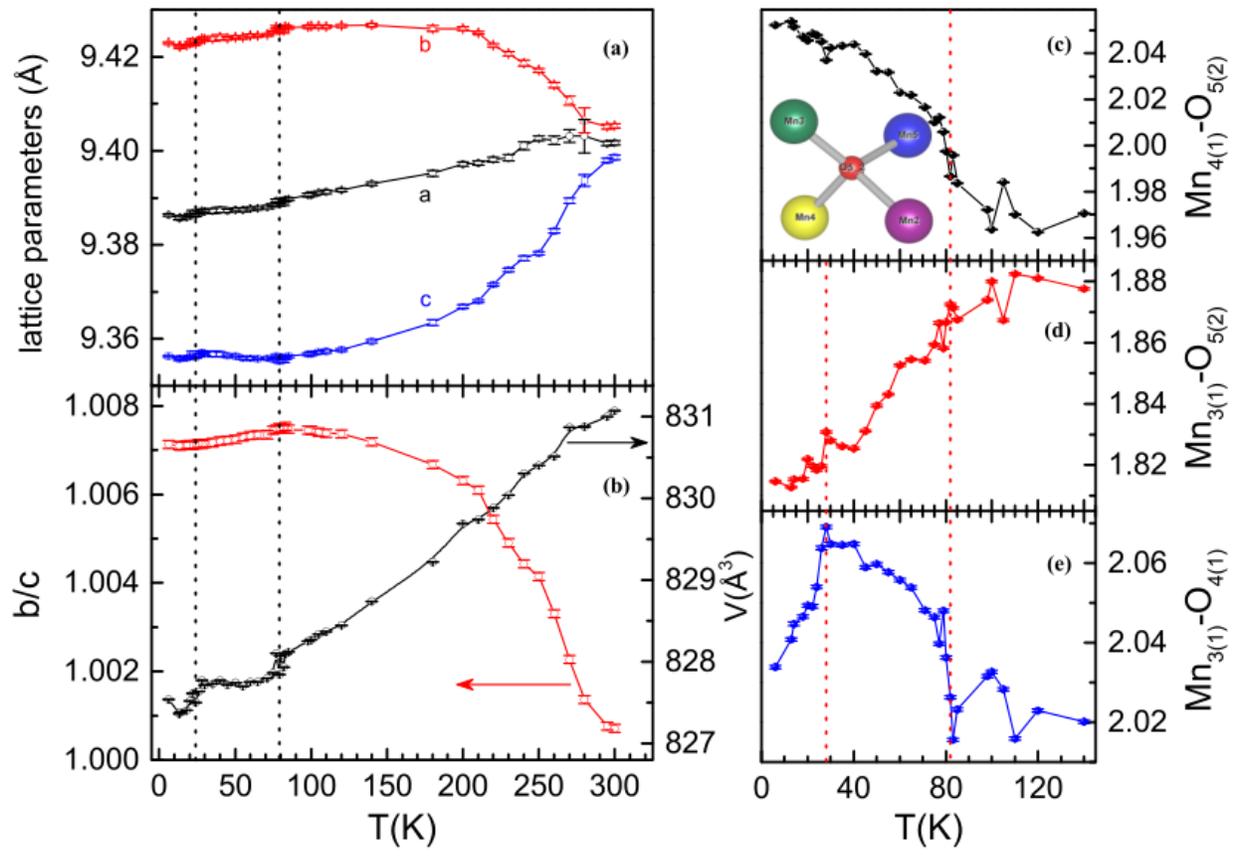

**Fig. 6**